\documentclass{ws-procs9x6}            
\begin{document}
\title{Three body open flavor decays of higher vector charmonium and bottomonium}
\author{Xin-Zhen~Weng$^{1,*}$,
Li-Ye~Xiao$^{2,\dagger}$,\\
Wei-Zhen~Deng$^{1,\ddagger}$,
Xiao-Lin~Chen$^{1,\S}$,
and
Shi-Lin~Zhu$^{1,3,4,\|}$}
\address{$^{1}$School of Physics and State Key Laboratory of Nuclear Physics and Technology, Peking University, Beijing 100871, China\\
$^{2}$University of Science and Technology Beijing, Beijing 100083, China\\
$^{3}$Center of High Energy Physics, Peking University, Beijing 100871, China\\
$^{4}$Collaborative Innovation Center of Quantum Matter, Beijing 100871, China\\
E-mail: $^{*}$xzhweng@pku.edu.cn
$^{\dagger}$lyxiao@ustb.edu.cn\\
$^{\ddagger}$dwz@pku.edu.cn
$^{\S}$chenxl@pku.edu.cn
$^{\|}$zhusl@pku.edu.cn}
\begin{abstract}
With an extended quark pair creation model we systematically study the OZI-allowed three body open flavor decays of higher vector charmonium and bottomonium states.
We obtain that the $BB^*\pi$ and $B^*B^*\pi$ partial decay widths of $\Upsilon(10860)$ are consistent with experiment, and the corresponding partial decay widths of $\Upsilon(11020)$ can reach up to 2$\sim$3 MeV.
Meanwhile the partial widths of $DD^*\pi$ and $D^*D^*\pi$ modes for most higher vector charmonium states can reach up to several MeV.
\end{abstract}
\bodymatter

\section{Introduction}

After the observation of $X(3872)$ by the Belle Collaboration in 2003~\cite{Choi:2003ue}, lots of charmoniumlike states have been found in experiment~\cite{Chen:2016qju}.
Among them, the $1^{--}$ states are of special importance because they can be easily produced in the $e^+e^-$ annihilation.
To reveal the nature of charmoniumlike states, it is important to have a better understanding of the normal charmonium states. Considering the strong decay properties playing a pivotal role in uncovering the nature of hadrons, in the present work we focus on the strong decays of charmonium.   
Besides the two body open flavor decays, three body open flavor decays are also important to the study of charmonium states with the experimental process.
In Ref.~\citenum{Xiao:2018iez}, we extended the quark pair creation(${^3P_0}$) model to second order and studied the $\psi(4660){\to}\Lambda_{c}\bar{\Lambda}_{c}$ process.
In this work, we further use the extended ${^3P_0}$ model to study the three body open flavor decays of higher vector charmonium and bottomonium states.
%

\section{The ${^3P_0}$ Model}
\label{Sec:Model}

In the ${^3P_0}$ model, a light quark pair is created from the vacuum, and then regroups with the quarks within the initial hadron to produce two daughter hadrons~\cite{Micu:1968mk,LeYaouanc:1972vsx,LeYaouanc:1973ldf}.
The corresponding interaction Hamiltonian reads~\cite{Ackleh:1996yt}
\begin{equation}\label{eqn:3P0:interaction}
H_{q\bar{q}} =
\gamma\sum_{f}2m_{f}\int\mathrm{d}^{3}{x}\bar{\psi}_{f}\psi_{f},
\end{equation}
where $\gamma$ is the quark pair creation strength parameter, and $\psi_{f}$ stands for the Dirac field of light quark $f$ with constituent quark mass $m_{f}$.
\begin{figure}
\begin{center}
\includegraphics[width=2.5in]{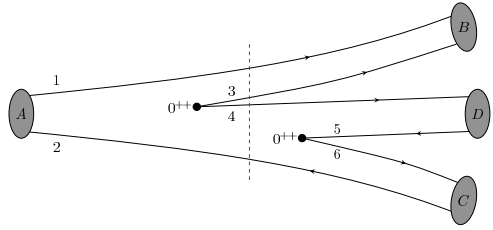}
\end{center}
\caption{The quarkonium ($A$) decays into three mesons ($B+C+D$).
The intermediate state is marked by a red dashed line.}
\label{fig:QPC_M3M}
\end{figure}
To study the decays of higher heavy quarkonium into two light-heavy mesons plus a light meson (see Fig.~\ref{fig:QPC_M3M}), we extend the ${^3P_0}$ model to second order, where two light $q\bar{q}$ pairs are created from the vacuum~\cite{Xiao:2018iez,Weng:2018ebv}.
The corresponding helicity amplitude  $\mathcal{M}^{M_{J_A}M_{J_B}M_{J_C}M_{J_D}}$ is defined as
\begin{eqnarray}\label{eqn:amplitude:definition}
\delta^{3}\left(\mathbf{p}_{A}-\mathbf{p}_{B}-\mathbf{p}_{C}-\mathbf{p}_{D}\right)
\mathcal{M}^{M_{J_A}M_{J_B}M_{J_C}M_{J_D}}
\nonumber\\
=
\sum_{k}
\frac{\left\langle{BCD}|H_{q\bar{q}}|{k}\right\rangle\left\langle{k}|H_{q\bar{q}}|{A}\right\rangle}{E_k-E_A}
\end{eqnarray}
where $|k\rangle$ is the intermediate state; $E_{A}$ ($E_{k}$) is energy of the initial (intermediate) state; $\mathbf{p}_{i}$'s stand for the momenta of the hadrons.
We apply the quark-hadron duality~\cite{Shifman:2000jv} %
to simplify the calculation.
Since the intermediate state differs from the initial state by a created quark pair at the quark level, we assume~\cite{Xiao:2018iez,Weng:2018ebv}
\begin{equation}
E_{k}-E_{A}\approx2m_{q}.
\end{equation}
Under this approximation, we can rewrite the Eq.~(\ref{eqn:amplitude:definition}) as
\begin{eqnarray}\label{eqn:amplitude:definition:rewrite}
\delta^{3}\left(\mathbf{p}_{A}-\mathbf{p}_{B}-\mathbf{p}_{C}-\mathbf{p}_{D}\right)
\mathcal{M}^{M_{J_A}M_{J_B}M_{J_C}M_{J_D}}
\approx
\frac{\left\langle{BCD}|H_{q\bar{q}}H_{q\bar{q}}|{A}\right\rangle}{2m_{q}}.
\end{eqnarray}
It should be mentioned that the above assumption is reasonable at the hadron level as well. 
Taking charmonium for example, the possible intermediate states with quantum number $J^{PC}=1^{--}$ are $D\bar{D}_1$, $D^*\bar{D}_0$, $D^*\bar{D}_1$, $D^*\bar{D}_2$, ${J/\psi}f_0(500)$, $h_c(1P)\pi$, $h_c(1P)\eta$, $\chi_{c0}(1P)\omega$, $\chi_{c2}(1P)\omega$ and tetraquark states~\cite{Cui:2006mp} and so on.
Their masses are approximately $(4.0\sim4.1)~\text{GeV}$. Thus it is justness to take $E_{k}-E_{A}$ as a constant for the higher mass states like $\psi(4360)$, $\psi(4415)$, and $\psi(4660)$.
But for the lower mass state, such as $\psi(4040)$ or $\psi(4160)$, the value of $E_{k}-E_{A}$ is sensitive to the masses of intermediate state due to its smallness.
In this case, taking $E_{k}-E_{A}$ as a constant will result in large uncertainties in the calculation.
We will focus on the higher mass states in the following.
%

\section{Results and Discussions}
\label{Sec:Results}

For the higher vector charmonium states, we study the decay properties of $\psi(4360)$, $\psi(4415)$, and $\psi(4660)$. The theoretical results are collected in Tables~\ref{table:width:cc:4380:DDpi}--\ref{table:width:cc:4660:DDpi}. It is shown that taking $\psi(4360)$ as $\psi(4{^3S_1})$~\cite{Segovia:2013wma} or $\psi(3{^3D_1})$~\cite{Li:2009zu}, the $DD^*\pi$ partial decay width can reach up to $1~\text{MeV}$. The partial decay width is large enough to be observed in future experiments if $\psi(4360)$ indeed turns out to be the charmonium state $\psi(4{^3S_1})$ or $\psi(3{^3D_1})$.
\begin{table}
\begin{minipage}[b]{0.37\textwidth}
\tbl{Partial decay widths (MeV) of the vector charmonium with a mass of 4368~MeV.}
{\begin{tabular}{@{}lccc@{}}
		\toprule
		State&$\psi(4{^3S_1})$&$\psi(3{^3D_1})$ \\
		\colrule
		$\Gamma_{DD\pi}$ & $0.27$ & $0.14$ \\
		$\Gamma_{DD^{*}\pi}$ & $1.40$ & $1.21$ \\
		$\Gamma_{D^{*}D^{*}\pi}$ & $0.60$ & $0.25$ \\
		$\Gamma_{DD\eta}$ & $0.6$~keV & $0.3$~keV \\
		\botrule
\end{tabular}}
\label{table:width:cc:4380:DDpi}
\end{minipage}~~~~
\begin{minipage}[b]{0.6\textwidth}
\tbl{Partial decay widths (MeV) of the vector charmonium with a mass of $4421~\text{MeV}$.}
{\begin{tabular}{@{}lcccc@{}}
		\toprule
		State&$\psi\left(4{^3S_1}\right)$&$\psi\left(5{^3S_1}\right)$&$\psi\left(3{^3D_1}\right)$ \\
		\colrule
		$\Gamma_{DD\pi}$& $0.38$& $0.11$& $0.21$ \\
		$\Gamma_{DD^{*}\pi}$& $2.01$& $0.96$ & $1.84$ \\
		$\Gamma_{D^{*}D^{*}\pi}$& $1.07$ & $0.59$ & $0.52$ \\
		$\Gamma_{DD\eta}$& $5.4\text{ keV}$ & $1.7\text{ keV}$& $2.9\text{ keV}$ \\
		\botrule
\end{tabular}}
\label{table:width:cc:4415:DDpi}
\end{minipage}
\end{table}

For the $\psi(4415)$ state, when we take it as the $\psi(4{^3S_1})$ assignment, the decay widths of the $DD^*\pi$ and $D^*D^*\pi$ modes are larger than $1~\text{MeV}$, while the $DD\pi$ mode is also sizable, as shown in Table~\ref{table:width:cc:4415:DDpi}.
The predicted branching ratios of the $DD\pi$ and $DD^*\pi$ channels are consistent with the upper limits ($2.2\%$ and $11\%$ respectively) measured by the Belle Collaboration~\cite{Pakhlova:2007fq,Pakhlova:2009jv}.
We also consider the possibility of $\psi(5{^3S_1})$ or $\psi(3{^3D_1})$, which gives similar results.
More experiments are needed to determine the inner structure of the $\psi(4415)$ state.

As to $\psi(4660)$ we have studied its decays into $\Lambda_{c}\bar{\Lambda}_{c}$ in Ref.~\citenum{Xiao:2018iez}.
This state has many possible assignments, namely $\psi(4S,5S,6S)$ and $\psi(3D,4D,5D)$~\cite{Xiao:2018iez}.
With those possible assignments, we calculate its decay properties, as list in Table~\ref{table:width:cc:4660:DDpi}.
From Table~\ref{table:width:cc:4660:DDpi}, we see that the $DD^*\pi$ and $D^*D^*\pi$ channels are always the dominant channels, and the corresponding widths can reach up to several MeV.
The decay widths of $DD\rho$ and $DD\omega$ modes are also quite large.
Moreover, if $\psi(4660)$ is $\psi(3D)$ state, the $DD\rho$ decay width can reach up to $1.86~\text{MeV}$.
\begin{table}
\tbl{The partial decay width (MeV) of the vector charmonium with a mass of $4643~\text{MeV}$.}
{\begin{tabular}{@{}lccccccccc@{}}
\toprule
State&$\psi(4{^3S_1})$&$\psi(5{^3S_1})$&$\psi(6{^3S_1})$&$\psi(3{^3D_1})$&$\psi(4{^3D_1})$&$\psi(5{^3D_1})$ \\
\colrule
$\Gamma_{DD\pi}$& $1.14$& $0.31$& $0.09$& $0.63$& $0.17$& $0.05$ \\
$\Gamma_{DD^*\pi}$& $6.65$& $2.83$& $1.10$& $6.99$&$2.99$& $1.16$ \\
$\Gamma_{D^*D^*\pi}$& $5.97$& $2.68$& $1.13$ & $4.12$ & $2.11$ & $0.96$ \\
$\Gamma_{DD\rho}$ & $0.85$ & $0.41$ & $0.16$ & $1.86$ & $0.64$ & $0.22$ \\
$\Gamma_{DD\omega}$ & $0.24$ & $0.12$ & $0.05$ & $0.59$ & $0.20$ & $0.07$ \\
$\Gamma_{DD\eta}$ & $53.2~\text{keV}$ & $15.3~\text{keV}$ & $4.2~\text{keV}$ & $29.1~\text{keV}$ & $8.2~\text{keV}$ & $2.2~\text{keV}$ \\
$\Gamma_{DD^*\eta}$ & $0.25$ & $0.12$ & $0.05$ & $0.20$ & $0.11$ & $0.05$ \\
$\Gamma_{D^*D^*\eta}$ & $58.2~\text{keV}$ & $38.7~\text{keV}$ & $19.1~\text{keV}$ & $8.5~\text{keV}$ & $9.9~\text{keV}$ & $7.5~\text{keV}$ \\
$\Gamma_{D_{s}D_{s}\eta}$ & $3.0~\text{keV}$ & $0.8~\text{keV}$ & $0.2~\text{keV}$ & $1.6~\text{keV}$ & $0.4~\text{keV}$ & $0.1~\text{keV}$ \\
$\Gamma_{D_{s}D_{s}^*\eta}$ & $1.9~\text{keV}$ & $1.3~\text{keV}$ & $0.6~\text{keV}$ & $12~\text{eV}$ & $11~\text{eV}$ & $7~\text{eV}$ \\
\botrule
\end{tabular}}
\label{table:width:cc:4660:DDpi}
\end{table}

Besides the vector charmonium states, we also discuss the decay properties of $\Upsilon(10860)$ and $\Upsilon(11020)$, which are above the $BB\pi$ threshold. The predictions are presented in Table~\ref{table:width:bb:BBpi}.
For $\Upsilon(10860)$, the partial decay ratios of the $BB\pi$ and $B^*B^*\pi$ modes are consistent with the experimental data, and the partial ratio of the $BB^*\pi$ mode is also very close to the experimental data~\cite{Tanabashi:2018oca}.
For $\Upsilon(11020)$, the decay widths of $BB\pi$, $BB^*\pi$, and $B^*B^*\pi$ are $0.34$, $3.17$ and $2.69~\text{MeV}$, respectively.
The latter two are quite large, and have a good potential to be observed by the Belle II Collaboration in the near future.
\begin{table}
\tbl{The $B^{(*)}\bar{B}^{(*)}\pi$ partial decay widths of the vector bottomonium (MeV). $\mathcal{B}_{\text{exp}}$ represents the branching ratio for each corresponding channel.}
{\begin{tabular}{@{}cccccccccc@{}}\toprule
Meson & State & Mode & $\Gamma_{\text{th}}$ & $\mathcal{B}_i$ & $\Gamma_{\text{th.}}$~\cite{Simonov:2008cr} & $\mathcal{B}_{\text{exp.}}$~\cite{Tanabashi:2018oca} \\
\colrule
$\Upsilon(10860)$&$5{^3S_1}$&$BB\pi$&$0.20$&$0.4\%$&&$(0.0\pm1.2)\%$ \\
&&$BB^*\pi$&$1.22$&$2.4\%$&(23--30)~\text{keV}&$(7.3\pm2.3)\%$ \\
&&$B^*B^*\pi$~&$0.61$&$1.2\%$&(5--6.6)~\text{keV}&$(1.0\pm1.4)\%$ \\
$\Upsilon(11020)$&$6{^3S_1}$&$BB\pi$&$0.34$&$0.7\%$ \\
&&$BB^*\pi$&$3.17$&$6.5\%$& \\
&&$B^*B^*\pi$&$2.69$&$5.5\%$& \\
\botrule
\end{tabular}}
\label{table:width:bb:BBpi}
\end{table}
%

\section*{Acknowledgments}

This project is supported by the National Natural Science Foundation of China under Grants No. 11575008 and 11621131001 and National Key Basic Research Program of China (Grants No. 2015CB856700).
This project is also in part supported by China Postdoctoral Science Foundation under Grant No. 2017M620492.
%

\bibliographystyle{ws-procs9x6} 
\bibliography{myreference}

\begin{thebibliography}{10}

\bibitem{Choi:2003ue}
S.~K. Choi {\em et~al.}, {\em Phys. Rev. Lett.} {\bf 91}, p. 262001  (2003).

\bibitem{Chen:2016qju}
H.-X. Chen, W.~Chen, X.~Liu and S.-L. Zhu, {\em Phys. Rept.} {\bf 639}, 1
  (2016).

\bibitem{Xiao:2018iez}
L.-Y. Xiao, X.-Z. Weng, Q.-F. L{\"u}, X.-H. Zhong and S.-L. Zhu, {\em Eur.
  Phys. J.} {\bf C78}, p. 605  (2018).

\bibitem{Micu:1968mk}
L.~Micu, {\em Nucl. Phys. B} {\bf 10}, 521  (1969).

\bibitem{LeYaouanc:1972vsx}
A.~Le~Yaouanc, L.~Oliver, O.~Pene and J.~C. Raynal, {\em Phys. Rev.} {\bf D8},
  2223  (1973).

\bibitem{LeYaouanc:1973ldf}
A.~Le~Yaouanc, L.~Oliver, O.~Pene and J.~C. Raynal, {\em Phys. Rev.} {\bf D9},
  1415  (1974).

\bibitem{Ackleh:1996yt}
E.~S. Ackleh, T.~Barnes and E.~S. Swanson, {\em Phys. Rev.} {\bf D54}, 6811
  (1996).

\bibitem{Weng:2018ebv}
X.-Z. Weng, L.-Y. Xiao, W.-Z. Deng, X.-L. Chen and S.-L. Zhu, {\em Phys. Rev.}
  {\bf D99}, p. 094001  (2019).

\bibitem{Shifman:2000jv}
M.~A. Shifman, {Quark hadron duality}
\newblock arXiv:hep-ph/0009131.

\bibitem{Cui:2006mp}
Y.~Cui, X.-L. Chen, W.-Z. Deng and S.-L. Zhu, {\em HEPNP} {\bf 31}, 7  (2007).

\bibitem{Segovia:2013wma}
J.~Segovia, D.~R. Entem, F.~Fernandez and E.~Hernandez, {\em Int. J. Mod.
  Phys.} {\bf E22}, p. 1330026  (2013).

\bibitem{Li:2009zu}
B.-Q. Li and K.-T. Chao, {\em Phys. Rev.} {\bf D79}, p. 094004  (2009).

\bibitem{Pakhlova:2007fq}
G.~Pakhlova {\em et~al.}, {\em Phys. Rev. Lett.} {\bf 100}, p. 062001  (2008).

\bibitem{Pakhlova:2009jv}
G.~Pakhlova {\em et~al.}, {\em Phys. Rev.} {\bf D80}, p. 091101  (2009).

\bibitem{Tanabashi:2018oca}
M.~Tanabashi {\em et~al.}, {\em Phys. Rev.} {\bf D98}, p. 030001  (2018).

\bibitem{Simonov:2008cr}
{\relax Yu}.~A. Simonov and A.~I. Veselov, {\em JETP Lett.} {\bf 88}, 5
  (2008).

\end{thebibliography}
\end{document}